\documentclass[letterpaper, 12 pt]{article}  

\usepackage{graphics} 
\usepackage{epsfig} 
\usepackage{mathptmx} 
\usepackage{times} 
\usepackage{amsmath} 
\usepackage{amssymb}  
\usepackage{setspace}
\usepackage[margin=1in]{geometry}

\usepackage{color}
\usepackage{cite}
\usepackage{float}
\usepackage{ntheorem}

\usepackage{calrsfs}
\DeclareMathAlphabet{\pazocal}{OMS}{zplm}{m}{n}

\usepackage{subcaption}
\usepackage{cleveref}

\usepackage{algorithm,algcompatible,amsmath}
\algnewcommand\INPUT{\item[\textbf{Input:}]}%
\algnewcommand\OUTPUT{\item[\textbf{Output:}]}%

\usepackage{tikz}
\usetikzlibrary{arrows.meta,positioning}
\UseRawInputEncoding
\newtheorem{theorem}{Theorem}[section]

\newtheorem{assumption}{Assumption}[section]
\newtheorem{definition}{Definition}[section]

\newtheorem*{refproof}{Proof}

\title{\LARGE \bf
Reconfigurable Plug-and-play Distributed Model Predictive Control for Reference Tracking
}

\begin{document}

\onehalfspacing

\title{Reconfigurable Plug-and-play Distributed Model Predictive Control for Reference Tracking}
\author{Ahmed Aboudonia, Andrea Martinelli, Nicolas Hoischen and John Lygeros}
\date{}

\maketitle
\thispagestyle{empty}
\pagestyle{empty}

\begin{abstract}
    
    A plug-and-play model predictive control (PnP MPC) scheme is proposed for varying-topology networks to track piecewise constant references. The proposed scheme allows subsystems to occasionally join and leave the network while preserving asymptotic stability and recursive feasibility and comprises two main phases. In the redesign phase, passivity-based control is used to ensure that asymptotic stability of the network is preserved. In the transition phase, reconfigurable terminal ingredients are used to ensure that the distributed MPC problem is initially feasible after the PnP operation. The efficacy of the proposed scheme is evaluated by applying it to a network of mass-spring-damper systems and comparing it to a benchmark scheme. It is found that the novel redesign phase results in faster PnP operations, whereas the novel transition phase increases flexibility by accepting more requests. 
    
\end{abstract}

\let\thefootnote\relax\footnote{ Research supported by the Swiss National Science Foundation under NCCR Automation and by the European Research Council (ERC) under the European Union’s Horizon 2020 research and innovation programme grant agreement OCAL, No. 787845. (Corresponding Author: Ahmed Aboudonia)
}
\let\thefootnote\relax\footnote{The authors are with the Automatic Control Laboratory, Department of Electrical Engineering and Information Technology, ETH Zurich, 8092 Zurich, Switzerland. {\tt\small $\{$ahmedab,andremar,lygeros$\}$@control.ee.ethz.ch} and {\tt\small nhoischen@student.ethz.ch}}%

\section{INTRODUCTION}

Plug-and-play (PnP) control schemes have received considerable attention for the control of varying-topology networks, where agents occasionally join and leave the network 
\cite{stoustrup2009plug}. 
The need to integrate constraints has motivated research into combining Model Predictive Control (MPC) and PnP algorithms for ensuring stability and constraint satisfaction of such networks. 
In 
\cite{riverso2014plug}, 
decentralized PnP MPC schemes were developed applicable mainly to weakly coupled networks, as they consider all the dynamic coupling terms as disturbances.
For the regulation of strongly coupled networks, distributed PnP MPC schemes were developed in 
\cite{zeilinger2013plug,hu2018plug} 
based on a two-phase PnP algorithm. In the \textit{redesign phase}, once a PnP request is received, the MPC ingredients are computed for a specific set of subsystems to ensure stability of the new network. Then, the \textit{transition phase} computes a steady state from which the MPC problem is initially feasible for the new network. The PnP request is approved if the optimization problems of both phases are feasible.

We propose a novel PnP algorithm for a class of networked dynamical systems with Laplacian interconnections. Our scheme is inspired by \cite{zeilinger2013plug}, but addresses piecewise constant reference tracking problems, instead of regulation problems. Unlike \cite{zeilinger2013plug}, the proposed PnP algorithm updates the local terminal sets online based on the work in \cite{aboudonia2020distributed,aboudonia2021distributed,aboudonia2022online}. This yields larger feasible regions in the transition phase where more PnP requests can be accommodated and also avoids the computation of a global terminal set which might require a central coordinator. Furthermore, the redesign phase makes use of passivity theory to update the local terminal costs and controllers following the work in \cite{aboudonia2020passivity}. Hence, each subsystem whose terminal ingredients are updated solves only two optimization problems. This accelerates the PnP operation compared to \cite{zeilinger2013plug}, where an iterative procedure is used in the redesign phase. Unlike \cite{aboudonia2020passivity}, we consider a more general class of multiple-input-multiple-output systems whose coupling can be described with more than one Laplacian.

In Section II, we introduce the considered class of systems and recall the distributed MPC scheme developed in \cite{aboudonia2022online}. In Section III, we develop the novel PnP algorithm which runs once a PnP request is received. In Section IV, we evaluate the efficacy of the proposed scheme using a network of mass-spring-damper (MSD) systems followed by some concluding remarks in Section V.

\section{PROBLEM FORMULATION}

\label{pf}

We consider networked dynamical systems with a distributed structure which can be decomposed into a set $\pazocal{M}$ of $M$ subsystems. Two subsystems are assumed to be neighbours if they are dynamically coupled. The set of neighbours of the $i^\text{th}$ subsystem is denoted by $\pazocal{N}_i$ and is always assumed to include the $i^\text{th}$ subsystem. The notation $\pazocal{N}_i \backslash i$ excludes the $i^{\text{th}}$ subsystem from its set of neighbours. The $i^\text{th}$ subsystem is described using the linear time-invariant dynamics
\begin{equation}
    \label{sec2_dyn1}
    \begin{aligned}
        & x_i(t+1) = A_i x_i(t) + B_i u_i(t) + \sum\nolimits_{k=1}^{p_i} F_{i,k} v_{i,k}(t), \\
        & v_{i,k}(t) = \sum_{j \in \pazocal{N}_i \backslash i} l_{ij,k} (y_{j,k}(t)-y_{i,k}(t)), \ 
        y_{i,k}(t) = C_{i,k} x_i(t),
    \end{aligned}
\end{equation}
where $t \in \mathbb{Z}_+$ is the time index, $x_i \in \mathbb{R}^{n_i}$, $u_i \in \mathbb{R}^{m_i}$ and $y_i \in \mathbb{R}^{p_i}$ are the state, input and output vectors of the $i^\text{th}$ subsystem and $y_{i,k}$ refers to the $k^{\text{th}}$ output of the $i^{\text{th}}$ subsystem. The matrices $A_i$, $B_i$, vectors $F_{i,k}$, $C_{i,k}$ and scalars $l_{ij,k}$ are known and have appropriate dimensions. We assume that all subsystems have the same output dimension $p$ (i.e. $p=p_i$ for all $i \in \pazocal{M}$). We also assume that the interconnection due to the $k^{th}$ pair $(v_{i,k},y_{i,k})$ is described using the graph $\pazocal{G}_k$ with the Laplacian $L_k \in \mathbb{R}^{M \times M}$ whose entries are given by
\begin{equation}
    L_{ij,k} =
    \left\{
    \begin{matrix}
        \sum_{j \in \pazocal{N}_i \backslash i} l_{ij,k}, & i=j, \\
        -l_{ij,k}, & i \neq j, \ j \in \pazocal{N}_i, \\
        0, & i \neq j, \ j \notin \pazocal{N}_i, \\
    \end{matrix}
    \right.
\end{equation}
Note that if $j \in \pazocal{N}_i$, but the $k^{\text{th}}$ output of the $j^{\text{th}}$ subsystem does not affect the dynamics of the $i^{\text{th}}$ subsystem, then $l_{ij,k}$ is set to zero.
The dynamics \eqref{sec2_dyn1} can be also given by
\begin{equation}
	\label{sec2_dyn2}
	x_i(t+1) = A_{N_i} x_{N_i}(t) + B_i u_i(t), \quad
	y_i(t) = C_i x_i(t),
\end{equation}
where $x_{N_i} \in \mathbb{R}^{n_{N_i}}$ is a state vector comprising the states of the subsystems in the set $\pazocal{N}_i$ and $A_{N_i}$ can be constructed using $A_i$, $C_{i,k}$, $F_{i,k}$ and $l_{ij,k}$.
The states and inputs of the $i^\text{th}$ subsystem are constrained to polytopic sets,
\begin{equation}
	\begin{aligned}
		\label{sec2_cons}
		x_{i}(t) \in \pazocal{X}_{i} &= \{x_{i} \in \mathbb{R}^{n_{i}} : G_i x_{i} \leq g_i \}, \\
		u_i(t) \in \pazocal{U}_i &= \{u_i \in \mathbb{R}^{m_i} : H_i u_i \leq h_i \},
	\end{aligned}
\end{equation}
where the matrices $G_i \in \mathbb{R}^{q_i \times n_i}$, $H_i \in \mathbb{R}^{r_i \times m_i}$ and the vectors $g_i \in \mathbb{R}^{q_i}$, $h_i \in \mathbb{R}^{r_i}$ are known. Note that there are no constraints coupling states or inputs of different subsystems.

Assuming that the $i^\text{th}$ subsystem is required to track the target point $x_{r_i} \in \pazocal{X}_i$ whose corresponding input is $u_{r_i} \in \pazocal{U}_i$, we choose the cost function $ J_i = \sum_{t=0}^{T-1} \left\{||x_{i}(t)-{x}_{e_i}||_{Q_i} + ||u_i(t)-{u}_{e_i}||_{R_i} \right\} + ||x_i(T)-{x}_{e_i}||_{P_i} + ||{x}_{e_i}-x_{r_i}||_{S_i},$
where $T$ is the prediction horizon and $(x_{e_i},u_{e_i})$ is an artificial equilibrium of the $i^\text{th}$ subsystem. The cost function matrices $Q_i$, $R_i$, $P_i$ and $S_i$ are known positive definite matrices of appropriate dimensions and the artificial equilibrium of the $i^\text{th}$ subsystem is required to satisfy
\begin{equation}
	\label{sec2_eq}
	x_{e_i} = A_{N_i} x_{e_{N_i}} + B_i u_{e_i} \in \lambda_i \pazocal{X}_i, \quad
	u_{e_i} = K_i x_{e_i} + d_i \in \lambda_i \pazocal{U}_i,
\end{equation}
where $K_i \in \mathbb{R}^{m_i \times n_i}$ and $d_i \in \mathbb{R}^{m_i}$ define the terminal controller $\kappa_i(x_i)=K_i x_i + d_i$ and $\lambda_i \in (0,1)$. 
Note that $P_i$ and $K_i$ are computed offline such that $V(x)= \|x\|_P^2$ with $x=[x_1^\top,\hdots,x_M^\top]^\top$ and $P=\operatorname{diag}(P_1,\hdots,P_M)$ is a Lyapunov function for the overall system. 
Unlike the constraints \eqref{sec2_cons}, the costs of the subsystems are coupled indirectly through the dynamic coupling of the equilibrium encoded in \eqref{sec2_eq}.
The local variables $x_i$, $u_i$ and $x_{N_i}$ can be extracted from the global variables $x=[x_1^\top,\hdots,x_M^\top]^\top \in \mathbb{R}^{n \times n}$ and $u=[u_1^\top,\hdots,u_M^\top]^\top \in \mathbb{R}^{n \times n}$ through the projections
\begin{equation}
    \label{sec2_map}
	x_i = U_i x, \quad
	x_{N_i} = W_i x, \quad
	u_i = V_i u,
\end{equation}
where $U_i \in \{0,1\}^{n_i \times n}$, $W_i \in \{0,1\}^{n_{N_i} \times n}$ and $V_i \in \{0,1\}^{m_i \times m}$ are appropriately constructed.

To ensure asymptotic stability and recursive feasibility, the state $x_i(T)$ is constrained to lie in the ellipsoidal positively-invariant terminal set as follows
$ x_i \in \pazocal{X}_{f_i}=\{x_i \in \mathbb{R}^{n_i}:(x_i-c_i)^\top P_i (x_i-c_i)\leq \alpha_i^2\} $
where $\alpha_i$ and $c_i$ refer to the size and center of the local terminal set of the $i^\text{th}$ subsystem. This constraint can be approximated using the Schur Complement and diagonal dominance as 
\begin{equation}
    \label{sec2_ter}
    \begin{aligned}
        2 \alpha_i [P_i]_j \geq \bar{b}_{i_j} + \alpha_i \{P_i\}_j \ \forall j \in \{1,...,n_i\}, \\
        \alpha_i \geq \sum\nolimits_{j=1}^{n_i} \bar{b}_{i_j}, \quad -\bar{b}_{i_j} \leq (x_{i_j}-c_{i_j}) \leq \bar{b}_{i_j},
    \end{aligned}
\end{equation}
where $\bar{b}_i \in \mathbb{R}^{n_i}$ is a decision variable, $[\cdot]_j$ and $\{\cdot\}_j$ refer to the diagonal element and the summation of absolute values of the elements in the $j^{\text{th}}$ row of a matrix, respectively. While the terminal set shape represented by $P_i$ is computed offline, the size $\alpha_i$ and center $c_i$ are considered as decision variables computed online. To ensure the positive invariance of the local terminal sets, we use the constraint in \eqref{sec2_LMI1}  which is given overleaf in single column) and 
\begin{table*}
    \normalsize
    \begin{subequations}
    \label{sec2_LMI1}
    \begin{gather}
    \begin{split}
    	2 \alpha_i \left[ P_i^{-1} \right]_{k} 
    	\geq 
    	\alpha_i \left\{ P_i^{-1} \right\}_{k}
    	&+
    	\sum\nolimits_{l=1}^{n_{N_i}} \left( |A_{N_i} + B_i K_i U_i W_i^\top| \alpha_{N_i} \right)_{kl}
    	+
    	b_{i_k}
    	\ \forall k \in \{1,...,n_i\},
    	\\ 
    	&-b_i \leq (A_i+B_iK_i) c_{N_i} + B_i d_i - c_i \leq b_i,
    \end{split}
	\\
	\begin{split}
		2 \sum\nolimits_{j \in \pazocal{N}_i} \lambda_{ij} \left[ P_{ij} \right]_{k} 
		\geq 
		\sum\nolimits_{j \in \pazocal{N}_i} \lambda_{ij} \left\{ P_{ij} \right\}_{k}
		&+
		\sum\nolimits_{l=1}^{n_i} \left(|A_{N_i} + B_i K_i U_i W_i^\top| \alpha_{N_i}\right)^\top_{kl}
		\ \forall k \in \{1,...,n_{N_i}\},
		\\
		&\alpha_i - \sum\nolimits_{j \in \pazocal{N}_i} \lambda_{ij}
		\geq 
		\sum\nolimits_{l=1}^{n_i} b_{i_l},
	\end{split}
    \end{gather}
    \rule{\textwidth}{0.4pt}
    \vspace{-1cm}
    \end{subequations}
\end{table*}
\begin{equation}
    \label{sec2_LMI2}
    G_i^k c_i + \| G_i^k P_i^{-1/2} \|_2 \alpha_i \leq g_i^k,
\end{equation}
\begin{equation}
    \label{sec2_LMI3}
        H_i^l K_i c_i + H_i^l d_i + \| H_i^l K_i P_i^{-1/2} \|_2 \alpha_i \leq h_i^l,
\end{equation}
for all $i \in \pazocal{M}$, $k \in \{1,...,q_i\}$ and $l \in \{1,...,r_i\}$ where $P_{ij}=W_i U_j^\top P_j U_j W_i^\top$, $G_i^k$ and $H_i^l$ are the $k^{th}$ and $l^{th}$ rows of the matrices $G_i$ and $H_i$ respectively, $g_i^k$ and $h_i^l$ are the $k^{th}$ and $l^{th}$ entries of the vectors $g_i$ and $h_i$.
For the derivation of the constraints \eqref{sec2_LMI1}-\eqref{sec2_LMI3}, refer to \cite{aboudonia2022online}.
In summary, the online optimization problem is given by
\begin{equation}
    \label{sec2_ocp}
    \begin{aligned}
        min \sum_{i \in \pazocal{M}} J_i \ s.t. \ x_i(0) = x_{m_i}, \ 
        \eqref{sec2_dyn2}-\eqref{sec2_eq}, \ 
        \eqref{sec2_ter}-\eqref{sec2_LMI3} \\
        \forall i \in \{1,...,M\} \ \& \ t \in \{0,...,T\}.
    \end{aligned}
\end{equation}
where $x_{m_i}$ is the measured state of the $i^{\text{th}}$ subsystem and the decision variables are $x_i(t)$, $u_i(t)$, $x_{e_i}$, $u_{e_i}$, $\alpha_i$, $c_i$, $d_i$, $b_i$, $\bar{b}_i$ and $\rho_{ij}$ for all $t \in \{0,...,T\}$, $i \in \pazocal{M}$, $j \in \pazocal{N}_i$, $k \in \{1,...,q_i\}$ and $l \in \{1,...,r_i\}$.
    Although the distributed MPC problem \eqref{sec2_ocp} requires global information, it is still amenable to distributed optimization techniques such as consensus alternating direction method of multipliers (ADMM) and hence, can be solved in a distributed fashion \cite{boyd2011distributed}.

If subsystems leave or join, the network topology changes and hence the cost and constraints in \eqref{sec2_ocp} also change. Thus, the asymptotic stability of the closed-loop system and the recursive feasibility of the MPC problem are no longer guaranteed. To address this difficulty, we develop below a PnP algorithm which allows changes in the network topology while ensuring asymptotic stability and recursive feasibility.

\section{Plug-and-play Distributed MPC}

We discuss the PnP alogrithm to be run whenever a set $\pazocal{J}$ of new subsystems and/or a set $\pazocal{L}$ of existing subsystems send PnP requests to the network. After the $k^{\text{th}}$ PnP request, the subsystems of the $k^{\text{th}}$ network can be partitioned into three non-overlapping sets. These sets are the plugged-in set $\pazocal{J}$ comprising the new subsystems joining the network, the neighbour set $\pazocal{Z}$ including the subsystems belonging to neither $\pazocal{J}$ nor $\pazocal{L}$ and whose set of neighbours contains at least one subsystem in $\pazocal{J}$ or $\pazocal{L}$ and the non-neighbour set $\pazocal{O}$ comprising the subsystems belonging to neither $\pazocal{J}$ nor $\pazocal{Z}$.
Similar to \cite{zeilinger2013plug}, the proposed PnP algorithm comprises two phases: the redesign phase ensuring asymptotic stability of the $k^{\text{th}}$ network and the transition phase ensuring recursive feasibility of the corresponding MPC scheme. In the sequel, these two phases are discussed in detail.

\subsection{Redesign Phase}

In the redesign phase, the local terminal controllers of the plugged-in set $\pazocal{J}$ are designed. The controllers of the neighbour set $\pazocal{Z}$ are also modified to take into account the plugged-in set $\pazocal{J}$ and the plugged-out set $\pazocal{L}$. 
The newly-designed and modified terminal controllers are designed to ensure the asymptotic stability of the $k^{\text{th}}$ network in the absence of constraints. The weights of the local stage costs of a specific set of subsystems are then updated to ensure the asymptotic stability of the $k^{\text{th}}$ network in the presence of constraints. All parameters and variables mentioned in this section are those of the $k^{\text{th}}$ network unless otherwise stated.

First, we start with updating the terminal controllers of the subsystems in $\pazocal{J}$ and $\pazocal{Z}$ to asymptotically stabilize the $k^{\text{th}}$ network in the absence of constraints. The terminal dynamics of the subsystems in $\pazocal{J}$ and $\pazocal{Z}$ under the control law $u_i=K_ix_i+d_i$ is given by $x_i^+=(A_i+B_iK_i)x_i+\sum_{k=1}^{p_i} F_{i,k} v_{i,k}+B_id_i$. By appropriately shifting the equilibrium point, the dynamics can be expressed as 
\begin{equation}
\label{lemma1_1}
\Delta x_i^+=(A_i+B_iK_i)\Delta x_i + \sum\nolimits_{k=1}^{p_i} F_{i,k} \Delta v_{i,k}
\end{equation}
where $\Delta v_{i,k} = \sum_{j \in \pazocal{N}_i \backslash i} l_{ij,k}(\Delta y_{j,k} - \Delta y_{i,k})$ and $\Delta y_{i,k}=C_{i,k}\Delta x_{i,k}$.
In addition to the subsystem outputs $\Delta y_{i,k}$, to exploit passivity we define virtual outputs for each subsystem in $\pazocal{J}$ and $\pazocal{Z}$ as
\begin{equation}
    \label{sec3_vo}
    \Delta z_{i,k} = C_{i,k} \Delta x_i + D_{i,k} \Delta v_{i,k} \text{ for all } k \in \{1,...,p_i\}.
\end{equation}
where $D_i = \operatorname{diag}(D_{i,1},...,D_{i,p_i})$ is computed below. We also define $\Delta z_i = [\Delta z_{i,1}^\top,...,\Delta z_{i,p_i}^\top]^\top$, $\Delta v_i = [\Delta v_{i,1}^\top,...,\Delta v_{i,p_i}^\top]^\top$, $F_i = [F_{i,1},...,F_{i,p_i}]$ and $C_i=[C_{i,1}^\top,...,C_{i,p_i}^\top]$.

\begin{definition}[\hspace{-0.75pt}\cite{aliyu2011nonlinear}]
The dynamics in \eqref{lemma1_1} is strictly passive with respect to the pair $(\Delta v_i,\Delta z_i)$ if and only if there exist a storage function $V_i(\Delta x_i)\geq 0$ and a dissipation function $\gamma_i(\Delta x_i)\geq 0$ such that
\begin{equation}
    \label{sec3_pascond}
    V_i(\Delta x_i^+)-V_i(\Delta x_i) \leq \sum\nolimits_{k=1}^{p_i} \Delta v_{i,k} \Delta z_{i,k} - \gamma_i(\Delta x_i)
\end{equation}
\end{definition}
Note that we use virtual outputs because the dynamics in \eqref{sec2_dyn1} can not be passiviated using the actual outputs \cite{aboudonia2020passivity}.

Moreover, we define $L=\sum_{k=1}^{p} L_k \otimes e_k e_k^\top$ where $\otimes$ refers to the Kronecker product and $e_k \in \mathbb{R}^{p}$ is a unit vector whose $k^{\text{th}}$ element equals one. One can show that the matrix $L \geq 0$ since $L_k \geq 0$ and $e_k e_k^\top \geq 0$. Considering $C=\operatorname{diag}(C_1,...,C_M)$, we define $L_C=LC$ and $C_L=C^\top L^\top$ which can be decomposed as follows, $L_C=[L_{C_1}^\top,...,L_{C_M}^\top]^\top$ and $C_L=[C_{L_1}^\top,...,C_{L_M}^\top]^\top$ where $L_{C_i} \in \mathbb{R}^{p_i \times n_i}$ and $C_{L_i} \in \mathbb{R}^{n_i \times p_i}$. 
Recall that we denote the diagonal element in the $j^{\text{th}}$ row of a matrix by $[\cdot]_j$ and the summation of absolute values of the elements in the $j^{\text{th}}$ row by $\{\cdot\}_j$. Finally, we define the scalars $n_{ij}$ for all $i,j \in \{1,...,M\}$
such that $n_{ij} > 0$ if $i \in \pazocal{N}_j$, $n_{ij} = 0$ otherwise and $\sum_{j=1}^M n_{ij} \leq 1$ for all $i \in \{1,...,M\}$. We now make the following assumption on the parameters and variables of the zeroth network (before any PnP operation) which are different from those of the $k^{\text{th}}$ network used in the rest of this section; here $\mathbb{S}_{++}^n$ and $\mathbb{D}_{++}^n$ are the sets of $n \times n$ symmetric and diagonal positive definite matrices, respectively.

\begin{assumption}
    \label{ass1}
    Before receiving any PnP request, the dynamics of the $i^{\text{th}}$ subsystem in the zeroth network under the control law $u_i=K_i x_i + d_i$ is strictly passive with respect to the pair $(\Delta v_i, \Delta z_i)$ with quadratic storage function $V_i(\Delta x_i)=\Delta x_i^\top P_i \Delta x_i$ and dissipation function $\gamma_i(\Delta x_i)=\Delta x_i^\top \Gamma_i \Delta x_i$. The matrices $P_i \in \mathbb{S}^{n_i}_{++}$ $\Gamma_i \in \mathbb{D}^{n_i}_{++}$ and $D_i  \in \mathbb{D}^{n_i}_{++}$ satisfy $[\Gamma_i^{-1}]_j \leq \frac{1}{\{L_{C_i}\}_j+\epsilon_i}$ for all $j \in \{1,...,n_i\}$ and $[D_i]_j \leq \frac{1}{\{C_{L_i}\}_j}$ for all $j \in \{1,...,p_i\}$ such that $\{C_{L_i}\}_j > 0$ where $\epsilon_i$ are arbitrarily small positive scalars. Finally, the stage cost weights $Q_i \in \mathbb{S}^{n_i}_{++}$ and $R_i \in \mathbb{S}^{m_i}_{++}$ of the $i^{\text{th}}$ subsystem satisfy
    $
        \sum\nolimits_{j \in \pazocal{N}_i} n_{ij} W_i U_j^\top P_j U_j  W_i^\top - W_i U_i^\top (Q_i + K_i^\top R_i K_i) U_i W_i^\top
        - (A_{N_i}+B_i K_i U_i W_i^\top)^\top P_i (A_{N_i} + B_i K_i U_i W_i^\top) \geq 0.
    $
    \end{assumption}
Assumption \ref{ass1} can be ensured during the offline synthesis of the MPC scheme \eqref{sec2_ocp} for the zeroth network. This assumption ensures that the zeroth network is asymptotically stable in the presence of constraints under the MPC scheme \eqref{sec2_ocp} and in the absence of constraints under the controller $u_i=K_ix_i+d_i$ for all $i \in \pazocal{M}$ (see \cite{aboudonia2022online,aboudonia2020passivity} for more details).

\begin{theorem}
    \label{lemma1}
    After the $k^{\text{th}}$ PnP request, consider the $k^{\text{th}}$ network under the controller $u_i = K_i x_i +d _i$ for all $i \in \{1,...,M\}$. Under Assumption \ref{ass1}, this network is asymptotically stable in the absence of constraints if for each subsystem in $\pazocal{J}$ and $\pazocal{Z}$ of this network , there exist $ E_i \in \mathbb{S}_{++}^{n_i}$, $X_i \in \mathbb{D}_{++}^{n_i}$, $Y_i \in \mathbb{R}^{n_i \times m_i}$ and $D_i \in \mathbb{D}_{++}^{n_i}$ such that
    \begin{subequations}
    \label{sec3_MT0}
        \begin{equation}
        \label{sec3_MT3}
        \begin{bmatrix}
            E_i & \frac{1}{2} E_i C_i^\top & (A_i E_i + B_i Y_i)^\top & E_i \\
            \frac{1}{2} C_i E_i & D_i & {F}_i^\top & 0 \\
            (A_i E_i + B_i Y_i) & {F}_i & E_i & 0 \\
            E_i & 0 & 0 & X_i \\
        \end{bmatrix}
        \geq 0,
        \end{equation}
        \begin{equation}
        \label{sec3_MT4}
            [X_i]_j \leq \frac{1}{\{L_{C_i}\}_j+\epsilon_i} \forall j \in \{1,...,n_i\},
        \end{equation}
        \begin{equation}
        \label{sec3_MT5}
            [D_i]_j \leq \frac{1}{\{C_{L_i}\}_j} \forall j \in \{1,...,p_i\} \text{ such that } \{C_{L_i}\}_j > 0,
        \end{equation}
    \end{subequations}
    where $\epsilon_i$ are arbitrarily small positive scalars.
\end{theorem}
\begin{refproof}
    see Appendix.
\end{refproof}


Note that if the considered network is asymptotically stable under the controller $u_i=K_ix_i+d_i$ where $d_i = -K_i x_{r_i} + u_{r_i}$, then the network converges to the target point $x_{r_i}$. We now move to the second step of the redesign phase in which we update the costs of a specific set of subsystems to asymptotically stabilize the $k^{\text{th}}$ network in the presence of constraints.
For this purpose, we define the matrices $P=\operatorname{diag}(P_1,...,P_M)$, $K=\operatorname{diag}(K_1,...,K_M)$, $Q=\operatorname{diag}(Q_1,...,Q_M)$ and $R=\operatorname{diag}(R_1,...,R_M)$. 
To ensure the asymptotic stability of the $k^{\text{th}}$ network in the presence of constraints under the MPC controller \cite{aboudonia2021distributed}, the Lyapunov condition $V(x^+)-V(x) \leq -l(x)$ should be satisfied where $V(x)=x^\top P x$ and $l(x)=x^\top (Q+K^\top RK) x$. This, in turn, is satisfied if the inequality $P-(A+BK)^\top P(A+BK)-Q-K^\top R K \geq 0$. This condition can be ensured by appropriately tuning $Q$ and $R$. Although this inequality requires global information, it can be used to derive local conditions for each subsystem. 

\begin{theorem}
\label{prop4}
After the $k^{\text{th}}$ PnP request, consider the $k^{\text{th}}$ network under the MPC controller \eqref{sec2_ocp}. Under Assumption \ref{ass1}, this network is asymptotically stable in the presence of constraints if for each subsystem in $\pazocal{J}$, $\pazocal{Z}$ and their neighbours, there exist $Q_i \in \mathbb{S}_{++}^{n_i}$ and $R_i \in \mathbb{S}_{++}^{m_i}$ such that
    \begin{multline}
        \label{sec3A_MT3}
        \sum\nolimits_{j \in \pazocal{N}_i} n_{ij} W_i U_j^\top P_j U_j  W_i^\top - W_i U_i^\top (Q_i + K_i^\top R_i K_i) U_i W_i^\top \\
        - (A_{N_i}+B_i K_i U_i W_i^\top)^\top P_i (A_{N_i} + B_i K_i U_i W_i^\top) \geq 0.
    \end{multline}
\end{theorem}
\begin{refproof}
    see Appendix.
\end{refproof}

Notice that following \cite{aboudonia2021distributed}, asymptotic stability is ensured in Theorem \ref{prop4} with respect to the target point $x_{r_i}$.  Note also that each subsystem in $\pazocal{J}$, $\pazocal{Z}$, $\pazocal{L}$ and their neighbours can satisfy \eqref{sec3A_MT3} by solving a constrained optimization problem whose cost function is selected so that the $k^{\text{th}}$ network can meet the required performance specifications.
The PnP request is rejected if any of these optimization problems is infeasible. Note that the scalars $n_{ij}$ can be also considered as decision variables in such optimization problems and hence the PnP request is rejected if $\sum_{j=1}^M n_{ij} > 1$ for any $i \in \pazocal{M}$. 

\subsection{Transition Phase}

In the transition phase, we compute for the old network and the new subsystems steady states $x_{s_i}$ at which the PnP operations can take place. These steady states lead to safe PnP operations during which $x_i(t+1)=x_i(t)$ for all subsystems. To compute these steady states, an optimization problem is solved in a distributed manner. In this problem, we ensure that the old network and the new subsystems can reach their steady states starting from their current states before the PnP operation. We also ensure that the terminal sets of the new network and the plugged-out subsystems are reachable from the steady states after the PnP operation. Unlike \cite{zeilinger2013plug}, the terminal sets are considered as decision variable in this optimization problem. In the sequel, the superscript $(\cdot)^o$ refers to the parameters of the old network before the PnP operations, whereas the superscript $(\cdot)^n$ refers to those of the new network after the PnP operations.
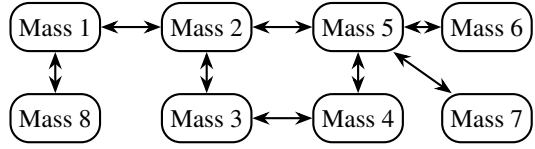
\begin{figure}
	\centering
	\scalebox{1}{\begin{tikzpicture}[thick,scale=0.8, every node/.style={scale=0.8}]
	\node[draw,thick,rectangle,rounded corners=0.25cm,minimum size=.8cm] (Mass1) {Mass 1};
	\node[draw,thick,rectangle,rounded corners=0.25cm,minimum size=.8cm, below = 0.55cm of Mass1] (Mass8) {Mass 8};
	\node[draw,thick,rectangle,rounded corners=0.25cm,minimum size=.8cm, right = 0.8cm of Mass1] (Mass2) {Mass 2};
	\node[draw,thick,rectangle,rounded corners=0.25cm,minimum size=.8cm, right = 0.8cm of Mass1, below = 0.55cm of Mass2] (Mass3) {Mass 3};
	\node[draw,thick,rectangle,rounded corners=0.25cm,minimum size=.8cm, right = 0.8cm of Mass2] (Mass5) {Mass 5};
	\node[draw,thick,rectangle,rounded corners=0.25cm,minimum size=.8cm, right = 0.8cm of Mass2, below = 0.55cm of Mass5] (Mass4) {Mass 4};
	\node[draw,thick,rectangle,rounded corners=0.25cm,minimum size=.8cm, right = 0.5cm of Mass5] (Mass6) {Mass 6};
	\node[draw,thick,rectangle,rounded corners=0.25cm,minimum size=.8cm, right = 0.8cm of Mass5, below = 0.55cm of Mass6] (Mass7) {Mass 7};
	\draw[Stealth-Stealth,thick] (Mass1) -- (Mass2);
	\draw[Stealth-Stealth,thick] (Mass1) -- (Mass8);
	\draw[Stealth-Stealth,thick] (Mass2) -- (Mass3);
	\draw[Stealth-Stealth,thick] (Mass2) -- (Mass5);
	\draw[Stealth-Stealth,thick] (Mass3) -- (Mass4);
	\draw[Stealth-Stealth,thick] (Mass4) -- (Mass5);
	\draw[Stealth-Stealth,thick] (Mass5) -- (Mass6);
	\draw[Stealth-Stealth,thick] (Mass5) -- (Mass7);
\end{tikzpicture}}
	\caption{Network Topology}
	\label{Topology}
\end{figure}

The prediction horizon in the transition phase is divided into two periods at $T_{pp} \in [0,T]$. In the first period, we drive the current state of the old network and the new subsystems to the steady state while satisfying the constraints before the PnP operation. For this purpose, we consider the dynamics and constraints of the old network and the new subsystems before joining the network as follows,
\begin{equation}
	\label{sec3_old}
	    \begin{aligned}
	        & \forall i \in \pazocal{M}^o \cup \pazocal{J} \\
	        & \forall t \in \{1,...,T_{pp}-1\}
	    \end{aligned}
        \left\{
	    \begin{aligned}
	        & x_i(0) = x_{m_i}, \quad x_i(T_{pp}) = x_{s_i} \\
	        & x_i(t+1) = A_{N_i}^o x_{N_i}(t) + B_i^o u_i(t), \\
	        & x_i(t) \in \pazocal{X}_i^o, \ u_i(t) \in \pazocal{U}_i^o \\
	    \end{aligned}
	    \right.
\end{equation}
In the second period, we drive the states of the new network and the plugged-out subsystems to their terminal sets while satisfying their dynamics and constraints after the PnP operation. For this purpose, we consider the dynamics and constraints of the new network and the plugged-out subsystems after leaving the network as follows,
\begin{equation}
	\label{sec3_new}
	    \begin{aligned}
	        & \forall i \in \pazocal{M}^n \cup \pazocal{L} \\
	        & \forall t \in \{T_{pp}+1,...,T\}
	    \end{aligned}
        \left\{
	    \begin{aligned}
	        & x_i(0) = x_{s_i}, \ \eqref{sec2_ter} \\
	        & x_i(t+1) = A_{N_i}^n x_{N_i}(t) + B_i^n u_i(t), \\
	        & x_i(t) \in \pazocal{X}_i^n, \ u_i(t) \in \pazocal{U}_i^n \\
	    \end{aligned}
	    \right.
\end{equation}
where \eqref{sec2_ter} considers the new network to ensure $x_i(T) \in \pazocal{X}_{f_i}^n$.

To guarantee that the steady states $x_{s_i}$ is a feasible equilibrium point for the old network and the new subsystems before joining the network, we consider the constraints,
\begin{equation}
	\label{pnp_trans2}
	\begin{aligned}
		{x_s}_i = A_i^o {x_s}_{N_i} + B_i^o {u_s}_i, \quad
		{x_s}_i \in \pazocal{X}_i^o, \quad
		u_i(T_{pp})={u_s}_i \in \pazocal{U}_i^o
	\end{aligned}
\end{equation}
where ${u_s}_i$ is the input corresponding to the steady state ${x_s}_i$ and $x_{s_{N_i}}$ is a concatenated vector which includes the steady states of the subsytems in the set $N_i^o$.
Finally, we also consider the artificial equilibrium constraints \eqref{sec2_eq} and the terminal set constraints \eqref{sec2_LMI1}-\eqref{sec2_LMI3} for the new network.
In summary, the transition phase is performed by solving the optimization problem
\begin{equation}
	\label{pnp_trans}
	min \sum\nolimits_{i \in \pazocal{M}^{o} \cup \pazocal{J}} ||x_{m_i}-x_{s_i}|| \text{ s.t. }
	\eqref{sec3_old}\text{-}\eqref{pnp_trans2}, \ \eqref{sec2_eq}, \  \eqref{sec2_LMI1}\text{-}\eqref{sec2_LMI3}, \ 
\end{equation}
where the dynamics and constraints of the new network are used in \eqref{sec2_eq} and \eqref{sec2_ter}-\eqref{sec2_LMI3}. If this optimization problem is feasible, the PnP request is accepted and the control sequence $[u_i(0),...,u_i(T_{pp)}]$ is applied to all subsystems before the PnP operation to drive the whole network to steady state. Once the steady state is reached, the distributed MPC problem \eqref{sec2_ocp} is solved recursively for the new network. Similar to the distributed MPC problem \eqref{sec2_ocp}, the transition phase can be solved in a distributed manner using ADMM \cite{boyd2011distributed}.

\begin{theorem}
    \label{theorem10}
	The MPC problem \eqref{sec2_ocp} is recursively feasible with the new network starting from the steady state $x_{s_i}$ if the optimization problem \eqref{pnp_trans} is feasible.
\end{theorem}
\begin{refproof}
    see Appendix.
\end{refproof}
\begin{figure}
	\centering
	\includegraphics[scale=0.275]{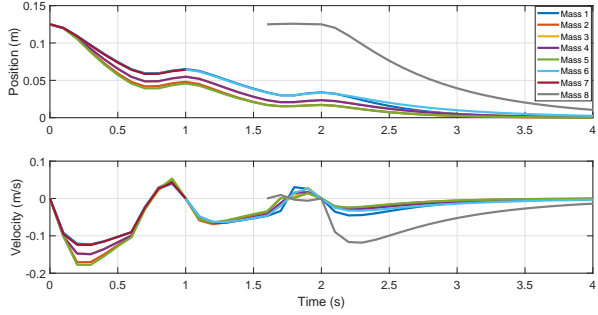}
	\caption{State trajectories of all masses}
	\label{Behavior}
\end{figure}

\section{Simulation Results}

The efficacy of the proposed PnP algorithm is illustrated using a network of MSD systems. The dynamics of each mass is given in continuous time by $\dot{x}_i=A_{ii}x_i+B_iu_i+
F_{i,1} l_{ij,1}(C_{j,1} x_j - C_{i,1} x_i) + F_{i,2} l_{ij,2}(C_{j,2} x_j - C_{i,2} x_i)$ 
where, for all $i \in \pazocal{M}$, $x_i$ comprises the position and velocity, $u_i$ is the applied force, $B_i=F_{i,1}=F_{i,2}= [0 \ 1]^\top$, $C_{i,1}=[1 \ 0]$, $C_{i,2}=[0 \ 1]$, $l_{ij,1}=k_{ij}$, $l_{ij,2}=c_{ij}$ and $A_{ii}=\begin{bmatrix} 0 & 1 \\ k_i/m_i & c_i/m_i \end{bmatrix}$. The parameters $m_i$, $k_i$ and $c_i$ represent the mass, stiffness and damping of mass $i$ and the parameters $k_{ij}$ and $c_{ij}$ represent the stiffness and damping connecting masses $i$ and $j$ where $m_i=k_i=1$, $c_i=2$, $k_{ij}=0.4$ and $c_{ij}=0.8$ for all $i \in \pazocal{M}$ and $j \in \pazocal{N}_i \backslash i$. All states and inputs are constrained between -1 and 1. The system is discretized using the method in \cite{riverso2013plug,aboudonia2020passivity} to preserve the structure. The sampling time and prediction horizon are  0.1 seconds and 8 timesteps, respectively. All optimization problems are solved using MATLAB with Yalmip \cite{Lofberg2004yalmip} and Mosek \cite{mosek} on a computer equipped with 
a 1.9-GHz Intel core i7-8550U processor.

The simulation considers a network of eight MSD systems with the topology in Fig. \ref{Topology}. The simulation starts without the eighth mass. The seventh mass sends a plug-out request at $t=0.6$ followed by a plug-in request by the eighth mass at $t=1.6$. 
Fig. \ref{Behavior} shows the state trajectories of all masses starting from the initial condition $[0.125 \ 0]^\top$ towards the target $[0 \ 0]^\top$. Note that the network is converging to the target despite the PnP operations. From $t=0.6$ till $t=1$ and from $t=1.6$ till $t=2$, all subsystems run the transition phase and move to their steady states at which their velocities are zero so that subsystem 7 is plugged out and subsystem 8 is plugged in, respectively. Apart from these two periods, all subsystems run the MPC scheme to converge to the target point. The redesign phase is run in parallel to the MPC scheme before the transition phase.

We compare our algorithm to the one in \cite{zeilinger2013plug}. First, we compare the redesign phase by running 100 simulations with different coupling strength $k_{18}=2.5r$ and $c_{18}=5r$ for all $r \in \{0.01,...,1\}$. We report the number of times $n_R$ the redesign phase is feasible and the mean $\mu_{R}$ and standard deviation $\sigma_{R}$ of the time taken in seconds by the redesign phase in Table \ref{table1}. Unlike the algorithm of \cite{zeilinger2013plug}, our algorithm fails for strong couplings, but is much faster. This is because the optimization problem of our algorithm is conservative, but completely decentralized, unlike that of \cite{zeilinger2013plug} that is distributed and solved using ADMM, requiring hundreds of iterations and hence, is computationally demanding. 
We also compare the transition phase by running 100 simulations with different initial position $p_8=0.5r$ for mass 8 for all $r \in \{0.01,...,1\}$. Both algorithms use ADMM to solve the optimization problem of the transition phase. We report the number of times $n_T$ the transition phase is feasible and the mean $\mu_{T}$ and standard deviation $\sigma_{T}$ of the time required in seconds per ADMM step. As shown in Table  \ref{table1}, our algorithm requires slightly more time as it has more constraints. 
But, it is feasible for all initial conditions due to the additional flexibility of computing the terminal sets online.

\section{Conclusions}

A PnP algorithm is developed for varying-topology networks tracking piecewise constant references. The algorithm comprises a redesign phase based on passivity-based control and a transition phase that makes use of reconfigurable terminal ingredients. The redesign phase results in faster PnP operations and the transition phase adds more flexibility by accepting more PnP requests compared to a benchmark scheme. Future work includes using clustering to steer a specific set of subsystems to steady state instead of steering the whole network during PnP operations.
\begin{table}[]
    \caption{The novel algorithm vs benchmark algorithm \cite{zeilinger2013plug}}
    \centering
    \begin{tabular}{|c|c|c|}
        \hline
        Redesign & Existing & New \\
        \hline
        $n_R$ & 100 & 82 \\
        \hline
        $\mu_R \ (s)$ & 29.73 & 0.008 \\
        \hline
        $\sigma_R \ (s)$ & 3.07 & 0.0006 \\
        \hline
    \end{tabular}
    \quad
    \begin{tabular}{|c|c|c|}
        \hline
        Transition & Existing & New \\
        \hline
        $n_T$ & 15 & 100 \\
        \hline
        $\mu_T \ (s)$ & 0.0024 & 0.003 \\
        \hline
        $\sigma_T \ (s)$ & 0.0004 & 0.0004 \\
        \hline
    \end{tabular}
    \label{table1}
\end{table}




\section*{APPENDIX}

{\bf Proof of Theorem \ref{lemma1}:}
    We use induction to prove the asymptotic stability of the $k^{\text{th}}$ network under the controller $u_i = K_i x_i +d_i$ for all $i \in \pazocal{M}$ in the absence of constraints. First, we start with the base case. Based on Assumption \ref{ass1}, the zeroth network before any PnP request is asymptotically stable. Next, we move to the inductive step. We assume that the  $(k-1)^{\text{th}}$ network is asymptotically stable (i.e. satisfies the conditions in Assumption \ref{ass1}). We derive conditions which ensure that the $k^{\text{th}}$ network is asymptotically stable.
    
    First, we derive conditions which ensure that all subsystems of the $k^{\text{th}}$ network are strictly passive, or equivalently that \eqref{sec3_pascond} holds for all $i \in \pazocal{M}$. Note that this condition is satisfied for $\pazocal{O}$ since the dynamics of such subsystems in the $k^{\text{th}}$ network remain the same as in the $(k-1)^{\text{th}}$ network which satisfies Assumption \ref{ass1}. Hence, it suffices to ensure that \eqref{sec3_pascond} holds for $\pazocal{J}$ and $\pazocal{Z}$. Following \cite{aboudonia2020passivity}, we take $V_i(\Delta x_i)=\Delta x_i^\top P_i \Delta x_i$ and $\gamma_i(\Delta x_i)=\Delta x_i^\top \Gamma_i \Delta x_i$, substitute \eqref{lemma1_1} and \eqref{sec3_vo} in \eqref{sec3_pascond}, use the Schur Complement twice to reach \eqref{sec3_MT3} where
    $E_i = P_i^{-1}, \quad Y_i = K_i P_i^{-1}, \quad X_i = \Gamma_i^{-1}$.
    The conditions $E_i \in \mathbb{S}_{++}^{n_i}$ and $X_i \in \mathbb{D}_{++}^{n_i}$ ensure that $V_i(x_i)\geq0$ and $\gamma_i(x_i)\geq0$.
    
    Second, we derive conditions which ensure that strict passivity of all subsystems in the $k^{\text{th}}$ network imply asymptotic stability of this network. For this, we define $\Delta z=[\Delta z_1^\top,...,\Delta z_M^\top]^\top$, $\Delta v=[\Delta v_1^\top,...,\Delta v_M^\top]^\top$,  $C=\operatorname{diag}( C_1,...,C_M)$ and $D=\operatorname{diag}(D_1,...,D_M)$. We also define the Lyapunov function $V(\Delta x)=\sum_{i=1}^M V_i(\Delta x_i)$ and the auxiliary function $\gamma(\Delta x)=\sum_{i=1}^M \gamma_i(\Delta x_i)=\Delta x^\top \Gamma \Delta x$ with $\Gamma = diag(\Gamma_1,...,\Gamma_M)$. Recall that \eqref{sec3_pascond} holds for all subsystems in $\pazocal{O}$ with quadratic storage and dissipation functions. Moreover, \eqref{sec3_MT3} implies that \eqref{sec3_pascond} holds for all subsystems in $\pazocal{J}$ and $\pazocal{Z}$ with quadratic storage and dissipation functions.
    By summing up \eqref{sec3_pascond} for all $i \in \pazocal{M}$, we reach $V(\Delta x^+)-V(\Delta x)\leq \Delta v^\top \Delta z - \gamma(\Delta x)$ where $\Delta z=C\Delta x+D\Delta v$, $\Delta v=-LC\Delta x$ and $\gamma(\Delta x)=\Delta x^\top \Gamma \Delta x$. 
    
    To  prove asymptotic stability of the $k^{\text{th}}$ network under the controller $u_i=K_ix_i+d_i$ for all $i \in \{1,...,M\}$ in the absence of constraints, it suffices to ensure that $\Delta v^\top \Delta z - \gamma(\Delta x) < 0$ or equivalently, $\Gamma+C^\top L C - C^\top L^\top D L C \geq \operatorname{diag}(\epsilon_1 I_{n_1},...\epsilon_M I_{n_M})$ where $\epsilon_i$ are arbitrarily small positive scalars and $I_{n_i}$ is an $n_i \times n_i$ identity matrix. Following \cite{aboudonia2020passivity}, we exploit the Schur Complement, diagonal dominance and the fact that $L \geq 0$, $X_i \in \mathbb{D}_{++}^{n_i}$ and $D_i \in \mathbb{D}_{++}^{n_i}$ to reach \eqref{sec3_MT4}-\eqref{sec3_MT5} for all subsystems in $\pazocal{J}$, $\pazocal{Z}$ and $\pazocal{O}$. Note that \eqref{sec3_MT4}-\eqref{sec3_MT5} are satisfied for all subsystems in $\pazocal{O}$ since the neighbours of these subsystems remain the same and hence, the matrices $C_{L_i}$ and $L_{C_i}$ do not change. Hence, it suffices to ensure that \eqref{sec3_MT4}-\eqref{sec3_MT5} are satisfied for all subsystems in $\pazocal{J}$ and $\pazocal{Z}$.
    
    If \eqref{sec3_MT0} is not satisfied for one or more subsystems, the PnP request is rejected. The network remains the same and, hence asymptotic stability is preserved and the conditions in Assumption \ref{ass1} hold until the next PnP request.
    If \eqref{sec3_MT0} is satisfied for all subsystems, the PnP request can be accepted depending on the next steps in the PnP algorithm. If it is, the controllers are redesigned based on \eqref{sec3_MT0} and the conditions in Assumption \ref{ass1} hold after the PnP operation is completed, hence until the next PnP request.

{\bf Proof of Theorem \ref{prop4}:}
    Using induction, we prove the asymptotic stability of the $k^{\text{th}}$ network under the MPC controller \eqref{sec2_ocp} in the presence of constraints. First, we start with the base case. Based on Assumption \ref{ass1}, the zeroth network before any PnP request is asymptotically stable. Next, we move to the inductive step. We assume that the  $(k-1)^{\text{th}}$ network is asymptotically stable (i.e. satisfies the conditions in Assumption \ref{ass1}). We derive conditions which ensure that the $k^{\text{th}}$ network is asymptotically stable.
    
    According to \cite{aboudonia2021distributed}, the $k^{\text{th}}$ network under \eqref{sec2_ocp} is asymptotically stable if the global inequality $P-(A+BK)^\top P(A+BK)-Q-K^\top R K \geq 0$ holds. This inequality holds if and only if $x^\top (P-(A+BK)^\top P(A+BK)-Q-K^\top R K) x \geq 0$ for all $x \in \mathbb{R}^n$. Due to the imposed structure on the matrices $P_i$, $K_i$, $Q_i$ and $R_i$, this inequality can be written as $\sum_{i=1}^M x_i^\top P_i x_i - \sum_{i=1}^M (A_{N_i}x_{N_i}+B_i K_i x_i)^\top P_i (A_{N_i}x_{N_i}+B_i K_i x_i) - \sum_{i=1}^M x_i^\top (Q_i+K_i^\top R_i K_i) x_i \geq 0$ or equivalently,
    \begin{multline*}
        \sum\nolimits_{i=1}^M \left(1-\sum\nolimits_{j=1}^M n_{ij}\right) x_i^\top P_i x_i + \sum\nolimits_{i=1}^M \left( \sum\nolimits_{j=1}^M n_{ij} \right) x_i^\top P_i x_i \\ 
        - \sum\nolimits_{i=1}^M (A_{N_i}x_{N_i}+B_i K_i x_i)^\top P_i (A_{N_i}x_{N_i}+B_i K_i x_i) \\ 
        - \sum\nolimits_{i=1}^M x_i^\top (Q_i+K_i^\top R_i K_i) x_i \geq 0.
    \end{multline*}
    
    Since $\sum_{j=1}^M n_{ij} \leq 1$ by design, it suffices to ensure that $\sum_{i=1}^M \left( \sum_{j=1}^M n_{ij} \right) x_i^\top P_i x_i - \sum_{i=1}^M (A_{N_i}x_{N_i}+B_i K_i x_i)^\top P_i (A_{N_i}x_{N_i}+B_i K_i x_i) - \sum_{i=1}^M x_i^\top (Q_i+K_i^\top R_i K_i) x_i \geq 0$.
    Using \eqref{sec2_map}, the resulting inequality can be reformulated as 
    \begin{multline*}
        \sum\nolimits_{i=1}^M \left( \sum\nolimits_{j=1}^M n_{ij} \right) x_{N_i}^\top W_i U_i^\top P_i U_i W_i^\top x_{N_i} \\ - \sum\nolimits_{i=1}^M x_{N_i}^\top (A_{N_i} +B_i K_i U_i W_i^\top)^\top P_i (A_{N_i} + B_i K_i U_i W_i^\top) x_{N_i}\\  - \sum\nolimits_{i=1}^M x_{N_i}^\top W_i U_i^\top (Q_i+K_i^\top R_i K_i) U_i W_i^\top x_{N_i} \geq 0.
    \end{multline*}
    
    One can show that this global inequality can be decomposed into the local inequalities,
    \begin{multline}
    	\sum_{j=1}^M n_{ij} x_{N_i}^\top W_i U_j^\top P_j U_j W_i^\top x_{N_i} - x_{N_i}^\top (A_{N_i} +B_i K_i U_i W_i^\top)^\top P_i (A_{N_i} + B_i K_i U_i W_i^\top) x_{N_i} \\
    	- x_{N_i}^\top W_i U_i^\top (Q_i+K_i^\top R_i K_i) U_i W_i^\top x_{N_i} \geq 0,
    \end{multline}
    for all $i \in \pazocal{M}$. Since $n_{ij}=0$ if $i$ and $j$ are not neighbours, this holds if and only if the local LMIs \eqref{sec3A_MT3} are satisfied for all $i \in \pazocal{M}$. Note that these LMIs are satisfied for all subsystems excluding $\pazocal{J}$, $\pazocal{Z}$ and their neighbours since the $(k-1)^{\text{th}}$ network satisfies Assumption \ref{ass1}. Hence, it suffices to ensure them for these subsystems.
    
    If \eqref{sec3A_MT3} is not satisfied for one or more subsystems, the PnP request is rejected. The network remains the same and, hence asymptotic stability is preserved and the conditions in Assumption \ref{ass1} hold until the $(k+1)^{\text{th}}$ PnP request.
    If \eqref{sec3A_MT3} is satisfied for all subsystems, the PnP request can be accepted depending on the next steps in the PnP algorithm. If it is, the stage cost weights are redesigned according to \eqref{sec3A_MT3} and the conditions in Assumption \ref{ass1} hold after the PnP operation is completed, hence until the next PnP request.

{\bf Proof of Theorem \ref{theorem10}:}
    Let $(x_i(0),...,x_i(T))$ and $(u_i(0),...,u_i(T-1))$ be feasible trajectories of the optimization problem \eqref{pnp_trans} for the $i^{\text{th}}$ subsystem. Therefore, each subsystem can be steered to its steady state $x_{s_i}$ in finite time since $x_i(T_{pp})=x_{s_i}$ and $u_i(T_{pp})=u_{s_i}$. Moreover, let $\alpha_i$ and $c_i$ refer to feasible size and center of the terminal set computed by \eqref{pnp_trans}.
    Define $x_{i,tail} = (x_{i,T+1},...,x_{i,T+T_{pp}})$ where 
    $x_{i,j} = (A_{N_i}^n+B_i^n K_i^n U_i^n W_i^{n^\top})^{j-T} x_{N_i}(T) + \sum_{k=1}^{j-T} (A_{N_i}^n+B_i^n K_i^n U_i^n W_i^{n^\top})^{k-1} B_i^n d_i$ and $u_{i,tail}=(K_i x_i(T)+d_i,K_i x_{i,T+1} + d_i,...,K_i x_{i,T+T_{pp}-1}+ d_i)$. Note that $x_{i,tail}$ is the evolution of the terminal dynamics and hence lies inside the terminal set described by $\alpha_i$ and $c_i$. Thus, $(x_i(T_{pp}),...,x_i(T),x_{i,tail})$, $(u_i(T_{pp}),...,u_i(T-1),u_{i,tail})$, $\alpha_i$ and $c_i$ represent a feasible solution for the MPC problem \eqref{sec2_ocp} applied to the new network. In other words, the MPC scheme for the new network is initially feasible. Recursive feasibility can thus be ensured using standard MPC arguments \cite{kouvaritakis2016model}.



\bibliographystyle{ieeetr}
\bibliography{root}

\end{document}